# Null hypothesis significance tests: A mix-up of two different theories – the basis for widespread confusion and numerous misinterpretations

Jesper W. Schneider

*Danish Centre for Studies in Research and Research Policy,*
*Department of Political Science & Government, Aarhus University,*
*Bartholins Allé 7, DK-8000, Aarhus C, Denmark*
[jws@cfa.au.dk](jws@cfa.au.dk)

**Abstract**

Null hypothesis statistical significance tests (NHST) are widely used in quantitative research in the empirical sciences including scientometrics. Nevertheless, since their introduction nearly a century ago significance tests have been controversial. Many researchers are not aware of the numerous criticisms raised against NHST. As practiced, NHST has been characterized as a 'null ritual' that is overused and too often misapplied and misinterpreted. NHST is in fact a patchwork of two fundamentally different classical statistical testing models, often blended with some wishful quasi-Bayesian interpretations. This is undoubtedly a major reason why NHST is very often misunderstood. But NHST also has intrinsic logical problems and the epistemic range of the information provided by such tests is much more limited than most researchers recognize. In this article we introduce to the scientometric community the theoretical origins of NHST, which is mostly absent from standard statistical textbooks, and we discuss some of the most prevalent problems relating to the practice of NHST and trace these problems back to the mix-up of the two different theoretical origins. Finally, we illustrate some of the misunderstandings with examples from the scientometric literature and bring forward some modest recommendations for a more sound practice in quantitative data analysis.

*"The statistician cannot excuse himself from the duty of getting his head clear on the principles of scientific inference, but equally no other thinking man can avoid a like obligation."*

(Fisher 1951, p. 2)

**Introduction**

To many researchers in the empirical sciences null hypothesis significance testing (NHST) is the primary epistemological doctrine used to organize and interpret quantitative research. NHST seemingly constitutes the sine qua non of 'objective' quantitative research. Nevertheless, since its introduction to the empirical sciences almost a century ago, statistical testing has caused much debate and controversy (for early critics, see e.g. Boring 1919; Berkson 1942; Rozeboom 1960). The criticism has gathered momentum through the decades in different fields with many articles arguing that such tests, if interpreted correctly, at best provide limited information, mostly indifferent for scientific progress, but worse, as Armstrong (2007) claims, such test may even harm scientific progress. Critics further argue that NHST as an inference model is based on invalid logic and that the procedure has severe methodological flaws. Perhaps most important, critics have incessantly pointed to the rote and often inane use of NHST, and how it is continuously misinterpreted and misapplied (for some general reviews, see e.g., Morrison & Henkel 1970; Oakes 1986; Gigerenzer 1993; Cohen 1994; Harlow et al. 1997, Nickerson 2000; Kline 2004; Ziliak & McCloskey 2008). Some defend NHST (e.g., Frick 1996; Abelson 1997; Cortina & Dunlap 1997; Chow 1998). Defenders claim that most failings of NHST are due to humans and that significance tests can play a role, albeit a limited one in research. Many critics will have none of this. To them history testifies that the so-called 'limited role' is not practicable, and if it were, the credulity was still strained due to the intrinsic flaws of such tests. Defenders typically suggest the need for better statistical education. Critics agree, but to them this is far from sufficient. Instead they point to the need for restricting the use of NHST and the general need for statistical reforms, such as focusing upon interval estimation (i.e., confidence intervals and effect sizes) (e.g., Kline 2004; Ellis 2010; Cumming 2012), or turning to likelihood or Bayesian statistics for inference and modelling (e.g., Royall 1997; Gill 2007). Critics also often stress that journal editorial policies must play a central role in such reforms. Nevertheless, despite mounting criticisms of NHST, significance testing continues to be (mis)used frequently; mindset and practice among researchers, reviewers and editors, seem hard to change.

The debate has hitherto been nearly absent from the scientometric literature; for an exception, see Schneider (2012; 2013), and replies from Bornmann and Leydesdorff (2013) and Leydesdorff (2013). A brief survey of recent volumes of the journal *Scientometrics* indicates that around 90-95 percent of the annual articles contain quantitative analyses and approximately half of them apply NHST. The use of NHST in the scientometric literature is somewhat smaller than other highly quantitative fields in the empirical sciences (e.g., Hubbard & Ryan 2000; Anderson Burnham & Thompson, 2000). However, an impressionistic glance of these recent articles, as well as bibliometric articles using NHST from other journals, suggest that bibliometric researchers, like numerous fellow researchers in other fields, also commonly invest these tests with far greater epistemic powers than they possess, resulting in misapplication and misinterpretation and undoubtedly also (unintentional) false knowledge claims (e.g., Ioannidis 2005; Gelman & Stern 2006).

In order to disclose some of the roots for the incessant misinterpretations in the research literature, we first discuss the two different theoretical approaches, which paradoxically have been anonymously merged into the widespread modern hybrid called 'null hypothesis significance testing' (i.e., NHST). Subsequently, we discuss the general confusion between *p* values and Type I error rates in NHST. Next we address some of the most prevalent confusions of NHST and persistent misinterpretations as a result. We exemplify some of these issues through recent examples from the scientometric literature. Finally, we point to some alternatives and suggest a list of recommendations for a better practice in quantitative data analysis.

## The origins of NHST: Fisher's 'significance tests' and Neyman-Pearson's 'hypothesis tests'

According to Gigerenzer (2004 p. 588), most NHST are performed as a 'null ritual' where a statistical null hypothesis of exactly 'zero association' or 'no difference' between population parameters are posited and tested mechanically. The actual research hypothesis usually corresponds to the alternative statistical hypothesis of a non-null effect, however, predictions of the research hypothesis or any alternative substantive hypotheses are usually not specified[1]. A conventional 'significance level' of 5%, often identified as *α*, is used for rejecting the null hypothesis. A probability measure, i.e. the *p* value is calculated and compared to *α*. A rigid decision process follows: if $p < α$ then the result is considered 'significant' otherwise not. If the result is 'significant' the research hypothesis is accepted. Results are reported as $p < .05$, $p < .01$, or $p < .001$ (whichever comes next to the obtained *p* value). This rote procedure is nearly always performed. According to Gigerenzer (2004), there are refined aspects, but this does not change the essence of the null ritual. To many researchers the null ritual is *the* 'objective' process leading to accurate inferences. In many respects, this is a misleading notion, as we will discuss in this article.

---

[1] Notice, other hypotheses to be nullified, such as a directional, non-zero or interval estimates are possible but seldom used, hence the 'null ritual'.

What is generally misunderstood is that what today is known, taught and practiced as NHST is actually an anonymous hybrid or mix-up of two divergent classical statistical theories, R. A. Fisher's 'significance test' and J. Neyman's and E. Pearson's 'hypothesis test' (e.g., Fisher 1925; 1935a; 1935b; 1935c; Neyman and Pearson 1928; 1933a; 1933b; Gigerenzer et al. 1989; Gigerenzer 1993). Even though NHST is presented somewhat differently in statistical textbooks (see Hubbard & Armstrong 2006), most of them do present *p* values, null hypotheses ($H_0$), alternative hypotheses ($H_A$), Type I (*α*) and II (*β*) error rates as well as statistical power, as if these concepts belong to one coherent theory of statistical inference, but this is not the case. Only null hypotheses and *p* values are present in Fisher's model. In Neyman-Pearson's model, *p* values are absent, but contrary to Fisher, two hypotheses are present, as well as Type I and II error rates and statistical power (Hubbard 2004).

Fisher argued that in a randomized experimental design, an observed result can be tested against a *single* statistical null hypothesis (Fisher 1925). The 'level of significance' or the measure of 'statistical significance' in Fisher's conception is the *p* value, a data-based random variable. The *p* value can be defined as $p = \Pr(T(X) \geq T(x)|H_0)$. Where, *p* is the probability of getting a test statistic $T(X)$ greater than or equal to the observed result, $T(x)$, as well as more extreme results, conditional on the null hypothesis of no effect or association being true, $H_0$ (Goodman 2008). Notice, it is a conditional probability of the observed test statistic as well as more extreme results of the test statistic which has not occurred. The *p* value is therefore a measure of (im)plausibility of observed as well as unobserved more extreme results, assuming a true null hypothesis. Fisher claimed that if the data are seen as being rare or highly discrepant under $H_0$, this would constitute inductive evidence against $H_0$, "… [e]ither an exceptionally rare chance has occurred, or the theory of random distribution [$H_0$] is not true [i.e., strong evidence against $H_0$]" (Fisher 1956, p. 39). This principle is also known as the law of improbability (See Royall 1997, p. 67). In Fisher's view, the *p* value is an epistemic measure of evidence from a single experiment and not a long-run error probability, and he also stressed that 'significance' depends strongly on the context of the experiment and whether prior knowledge about the phenomenon under study is available (Gigerenzer et al. 1989). To Fisher, a 'significant result provides evidence against $H_0$, whereas a non-significant result simply suspends judgment – nothing can be said about $H_0$.

Neyman and Pearson dismissed Fisher's 'significance test' and its inherent subjective interpretation, as well as the concept of 'inductive inference', as both mathematical inconsistent and a misconception of frequentist probability theory (Neyman and Pearson 1933a). They specifically rejected Fisher's quasi-Bayesian interpretation of the 'evidential' $p$ value, stressing that if we want to use only objective probability, we cannot infer from a single experiment anything about the truth of a hypothesis. For the latter we need subjective probabilities, a concept alien to frequentists like Fisher, Neyman and Pearson (Oakes 1986; Royall 1997). Neyman and Pearson argued that statistical inference can *only* be usefully applied to the problem of minimizing decision errors in the long-run (Neyman & Pearson 1933a; 1933b). Therefore they suggested 'hypothesis testing' of two complementary hypotheses, neither of which need to be a null hypothesis in Fisher's sense, but for simplicity we designate them $H_0$ and $H_A$, as a decision process in order to guide behavior. Neyman and Pearson argued that one could not consider a null hypothesis unless one could conceive at least one plausible alternative hypothesis. They therefore reasoned that two competing hypotheses (with two known probability density functions) invite a decision between two distinct courses of action, accepting $H_0$ or rejecting it in favor of $H_A$. Notice, 'accept' of H has nothing to do with the actual 'truth' of H. Neyman-Pearson's model is only about rules of behavior in the long-run, so that "we shall reject H when it is true not more than say, once in a hundred times, and in addition we may have evidence that we shall reject H sufficiently often when it is false" (Neyman & Pearson 1933a, p. 291). Consequently, Neyman-Pearson's model is concerned with error control, where $α$ is the probability of falsely rejecting $H_0$ under the assumption that it is true (Type I error), and $β$ is the probability of failing to reject $H_A$ when it is false (Type II error). Power[2], the complement of $β$ (1-$β$), is a loss function, whereby the most powerful test for a given significance level, sample and effect size should be pursued in advance of the study, to determine the long-run probability of accurately rejecting a false $H_0$ (Neyman & Pearson 1933a; 1933b). Accordingly, 'hypothesis tests' are concerned with minimizing Type II errors subject to a bound on Type I errors, and $α$ is a prescription for 'inductive behaviors' and *not* evidence for a specific result. Most importantly, error control is a pre-selected fixed measure; $α$ is therefore not a random variable based on the actual data, and $α$ applies only to infinitely random selections from the same finite population, not to an actual result in a single experiment. As a consequence, the fixed $α$ level implies that the decision process must be applied rigidly. If 5% is the desired long-run error rate, $H_0$ is rejected for an achieved significance level of 4.9% but accepted for an achieved significance level of 5.1%. Notice also that *p* values are absent in Neyman-Pearson's model, here decisions come from checking whether the test statistic is further than the critical value from the expected value of the test statistic. Table 1 below summarizes the different elements and concepts from Fisher's 'significance test' and Neyman-Pearson's 'hypothesis test'.

---

[2] Statistical power is the probability of rejecting $H_0$ when it is false (Cohen, 1988). Statistical power is affected by α and β levels, the size of the effect and the size of the sample used to detect it. These elements make it possible to define the probability density function for the alternative hypothesis.

**Table 1. Summary of the different elements and concepts in 'significance' and 'hypothesis' tests.**

| 'Significance test' (R. A. Fisher) | 'Hypothesis test' (J. Neyman and E. S. Pearson) |
|---|---|
| *p* value - a measure of the evidence against $H_0$ | $α$ and $β$ levels - provide rules to limit the proportion of decision errors |
| Calculated *a posteriori* from the observed data (random variable) | Fixed values, determined *a priori* at some specified level |
| Applies to any single experiment (short run) | Applies only to ongoing, identical repetitions of an experiment, not to any single experiment (long-run) |
| Roots in inductive philosophy: from particular to general | Roots in deductive philosophy: from general to particular |
| 'Inductive inference': guidelines for interpreting strength of evidence in data (subjective decisions) | 'Inductive behavior': guidelines for making decisions based on data (objective behavior) |
| Based on the concept of a 'hypothetical infinite population' | Based on a clearly defined population |
| Evidential, i.e., based on the evidence observed | Non-evidential, i.e., based on a rule of behavior |

We should emphasize that today the interpretation of research results most often ends with a verdict of 'statistical significance' or not. But it is important to recognize that both Fisher and Neyman-Pearson regarded their models as rather primitive tools to be handled with "discretion and understanding, and not as instruments which themselves give the final verdict" (Neyman & Pearson 1928, p. 232). Fisher regarded 'significance tests' as procedures to be used if one had only scant knowledge of the problem at hand and 'significant' results to Fisher meant they were worthy of notice and that they should be replicated in further experiments to gain more evidence.

**Early controversies: 'evidential measures' or 'error rates'**

Before becoming mixed-up in modern-day NHST, the scientific usefulness of error rates and the supposed evidential meaning of *p* values were already contested issues. It is well-known that Fisher and especially Neyman argued vehemently, sometimes acrimoniously, and that they never reconciled their opposing views (e.g., Gigerenzer et al. 1989). Neyman-Pearson's model is considered to be theoretically consistent and is generally accepted as 'frequentist orthodoxy' in mathematical statistics (e.g., Hacking 1965; Mayo 1996; Royall 1997). However, the price for theoretical clarity seems to be restricted utility in applied scientific work (e.g. Oakes 1986; Hurlbert & Lombardi 2009). The emphasis upon decision rules with stated error rates in infinitely repeated trials may be applicable to quality control in industrial settings, but seems less relevant to assessment of scientific hypotheses, as Fisher mockingly stressed (Fisher 1955).

Albeit Fisher originally suggested a 'significance level' of 5% for his 'significance tests', he later objected to Neyman-Pearson's dogmatic binary decision rules based on a predetermined $\alpha$ level, stressing that it was naïve for scientific purposes. Consequently, in later writings he argued that exact *p* values should be reported as evidence against $H_0$ without making hair-splitting rejection decisions (Fisher 1956).

On the other hand, the supposed 'objective' evidential nature of *p* values was also questioned early on, and Fisher's attempt of refutation of $H_0$ based on 'inductive inference' is generally considered to be logically flawed (e.g., Neyman & Pearson 1933a, Jeffreys 1961; Hacking 1965; Royall 1997, Chapter 3). Especially the fact that *p* values only test one hypothesis and are based on tail area probabilities was early on considered a serious deficiency (Jeffreys 1961). Dependence on tail area probabilities means that the calculation of *p* values is not only based on the observed results but also on 'more extreme results', i.e., results that have not occurred. In the words of Jeffreys' "[a] hypothesis that may be true may be rejected because it has not predicted … results which have not occurred" (Jeffreys 1939, VII, §7.2). This logical conundrum leads directly to practical problems, for example the so-called 'stopping rule' paradox, because what is 'more extreme results' depends on the actual sampling plan in a study (e.g., Wagenmakers 2007). The 'stopping-rule' paradox basically means that two studies, where we have the same number of cases, the same treatments and the same results, can have different *p* values, and thus perhaps different conclusions, simply because the researchers had different sampling plans in the two studies[3] (see Goodman 1999a for an illustrative example). A similar dependence on the sampling space was noticed by Berkson (1938), who showed that *p* values were depended on sample size. Consequently, *p* values can be equal in situations where the evidence is very different (i.e., different effect sizes in studies with different sample sizes), or different in situations where the evidence should be the same (i.e., same data in trials with different stopping rules). Sensitivity to 'stopping rules' and sample size are not properties for an inductive evidential measure (Good 1950; Royall 1997). According to Good (1950), an inductive measure of statistical evidence is defined as the relative support given to *two* hypotheses by the observed data. The law of likelihood states that it is a likelihood ratio of two hypotheses that measure evidence and that the sample space is irrelevant in that respect (see Royall 1997, p. 68). From this follows that *p* values are not inductive measures of evidence because their calculation involves only one hypothesis and is also based on unobserved data in the tail areas (e.g., Jeffreys 1939; Berkson 1942; Hacking 1965). Neyman and Pearson accordingly dismissed the notion of evidential *p* values and inductive inference and concentrated on error rates and behavior, but Fisher persisted that there was a link of some sort between *p* values and evidence, despite the violation of the likelihood principle – a principle Fisher himself in fact had developed substantially.

If the *p* value clearly violates the basic properties for an evidential measure, as Fisher very well knew, why did Fisher continue to treat *p* values as if they did have such evidential properties and why does the *p* value continue to be treated as a surrogate kind of evidential measure against $H_0$? Fisher never gave a satisfactory answer, but an indication could be the fact that the *p* value is often a monotonic function of the maximum likelihood ratio. Consequently, the *p* value is considered a transformation of the likelihood ratio, which compares the likelihood of $H_0$ to a *post hoc*, data-suggested hypothesis (i.e., the hypothesis with maximum likelihood versus $H_0$) (Goodman 2003). But according to Goodman (2003 p. 701), this is a measure that "violates a prime dictum of scientific research: to pre-specify hypotheses". A true measure of evidence uses a pre-specified alternative not dictated by the data (Goodman 2003). The all important question therefore is to what extent *p* values correlate empirically with 'evidence' and how precise are *p* values compared to true evidential measures? Using both likelihood and Bayesian methods, more recent research have demonstrated that *p* values overstate the evidence against $H_0$, especially in the interval between significance levels .01 and .05, and therefore can be highly misleading measures of evidence (e.g., Berger & Sellke 1987; Berger & Berry 1988; Goodman 1999a; Sellke et al. 2001; Hubbard & Lindsay 2008; Wetzels et al. 2011). What these studies show is that *p* values and true evidential measures only converge at very low *p* values. Goodman (1999a p. 1008) suggests that only *p* values less than .001 represent strong to very strong evidence against $H_0$. Berger and Sellke (1987) demonstrate that data yielding a *p* = .05 results in a posterior probability of at least 30% in support for $H_0$ for any objective prior distribution. Under somewhat different conditions, Goodman (1999a) show that when a result is 1.96 standard errors from its null value (i.e., *p* = .05), the minimum Bayes Factor is .15, meaning that $H_0$ gets 15% as much support as the best supported hypothesis. This is threefold higher than the *p* value of .05. The general conclusion from these studies is that the chance that rejection of $H_0$ is mistaken is far higher than is generally appreciated and this in turn, calls into question the validity of much published work based on comparatively small *p* values such as .05. Indeed, especially in clinical fields, false positive findings are of great concern and seems to be a major problem (e.g., Ioannidis (2005), and for an important correction to Ioannidis, see Goodman & Greenland (2007); for the problem of false-positives in psychology, see also Simmons et al. 2011).

---

[3] E.g., a sampling design where one either chooses to toss a coin until it produces a pre-specified pattern, or instead doing a pre-specified number of tosses. The results can be identical, but the *p* values will be different.

**More confusion: The amalgam of NHST**

Despite the contradictory elements in the two approaches and the implacable views between the contestants, from the 1940s onwards, applied textbooks in the social and behavioral sciences began to blend Fisher's and Neyman-Pearson's approaches into what today is known as NHST, and this usually without mentioning or citing its intellectual origins (Gigerenzer 1993; Dixon & O'Reilly 1999). Several authors have pointed out that this hybrid model would most certainly have been rejected by Fisher, Neyman and Pearson, albeit for different reasons (e.g., Gigerenzer 2004).

Within the NHST hybrid, Fisher's notion of $H_0$ is applied, but contrary to Fisher's doctrine, an alternative hypothesis is also included. Even so, the two hypotheses in NHST are not utilized according to Neyman-Pearson's decision model, where pre-specified error rates are used to find the most powerful test for the two complementary hypotheses. In NHST, non-significant results are sometimes treated as 'acceptance' of $H_0$, however, not as a consequence of Neyman-Pearson's decision rules based on the predetermined *α* and *β* levels. Remember, to Fisher non-significance meant suspension of judgment, not 'retaining', 'accepting' or 'failing to reject' $H_0$. NHST also pretends to select in advance Neyman-Pearson's fixed level of significance, *α*, but in practice ends up binning $p$ values into categories as strength of evidence against $H_0$, i.e., $p < \alpha$, and subsequently use these categories rigidly to establish whether results are statistical significant or not.

Blending these divergent concepts from two essentially incompatible approaches has for good reasons created massive confusion over the meaning of 'statistical significance' (Hubbard & Bayarri 2003). Several authors have noted that $p$ values are routinely misinterpreted as frequency-based 'observed' Type I error rates (i.e., an 'observed' $\alpha$), and at the same time are also used as evidential measures of evidence against $H_0$ (i.e., $p < \alpha$). While definitely confusing, it should be clarified that while $p$ values and error rates are both tail area probabilities derived in the same theoretical sampling distributions and both are frequently associated with a value of 5%, they are different entities. The Type I error rate, $\alpha$, is the probability of a set of *potential* outcomes that may fall anywhere in the tail area of the distribution under $H_0$. We cannot know beforehand which of these particular outcomes will occur (Goodman 1993). The tail area for the $p$ value is different, as it is known only after the result is observed and is based on a range of results under $H_0$ (i.e., the observed and more extreme ones). Thus, $p$ values do not estimate the conditional probability of a Type I error and $\alpha$ does not address specific evidence. As it is, $\alpha$ is an error probability specified before data are collected, and thus a property of the test with a long-run random sampling interpretation, $p$ is not (Hubbard & Armstrong 2006). As a result, the usual practice of reporting $p$ values in relation to a limited number of bins, i.e., $p < .05$, $p < .01$, $p < .001$, is problematic as it gives them the appearance of Type I error rates, this is known as 'roving alphas'(Goodman 1993 p. 489). As $\alpha$ must be fixed before the data collection, *ex post facto* attempts to reinterpret 'roving alphas' as variable Type I error rates is erroneous (Hubbard 2004). Further complicating matters, the $p$ value inequalities are at the same time also interpreted in an increasing evidential manner with labels such as significant ($p < .05$), highly significant ($p < .01$) and extremely significant ($p < .001$). Hubbard (2004) has referred to $p < \alpha$ as an 'alphabet soup', that blurs the distinctions between evidence ($p$) and error ($\alpha$), but the distinction is crucial as it reveals the basic differences underlying Fisher's ideas on 'significance testing' and 'inductive inference', and Neyman–Pearson views on 'hypothesis testing' and 'inductive behavior'. So complete is this misunderstanding over measures of evidence versus error rates that it is not viewed as even being a problem among influential institutions such as the APA Task Force on Statistical Inference (Wilkinson et al. 1999), and those writing the guidelines concerning statistical testing mandated in the APA Publication Manuals (Hubbard 2004). We may of course ask why researchers cannot report both $p$ and $\alpha$ in the same study. Hubbard and Bayarri (2003 p. 175) give the following answer: "[w]e have seen from a philosophical perspective that this is extremely problematic. We do not recommend it from a pragmatic point of view either, because the danger of interpreting the *p* value as a data dependent adjustable Type I error is too great, no matter the warnings to the contrary. Indeed, if a researcher is interested in the 'measure of evidence' provided by the *p* value, we see no use in also reporting the error probabilities, since they do not refer to any property that the *p* value has . . . Likewise, if the researcher is concerned with error probabilities, the specific *p* value is irrelevant".

Few have attempted to reconcile the two approaches theoretically (e.g., Lehmann 1993), but hitherto to no avail. Others have tried to reformulate them. Mayo (1996) with her 'error statistics' has proposed a reformulation of Neyman-Pearson's framework and recently Hurlbert and Lombardi (2009) have proposed a "neoFisherian" approach where they basically discard anything related to Neyman-Pearson concepts.

Finally, it is often overlooked that 'significance tests' as well as 'hypotheses tests' were specifically developed for controlled experimental settings, in Fisher's case agricultural research, and *not* studies based on observational data (Gigerenzer et al. 1989). Paramount to experimental settings and frequentist tests is randomization (i.e., random assignment and probability sampling) (e.g., Greenland 1990; Ludwig 2005). Randomization ensures statistical unbiasedness and provides a known probability distribution for the possible results under a specified hypothesis about the treatment effect. By capitalizing on what would happen in principle if repeated samples were generated independently by the same process, it then becomes possible to represent the uncertainty in the parameter estimates (Berk et al. 1995). Obviously, the notion of "repeated random sampling" where data come from an ongoing stream of independent, identically distributed (iid) values are crucial in this respect. Nature can produce an ongoing stream of iid values and in theory an infinite number of identical experimental trials can mirror this. However, much research in the social sciences is not experimental. Instead data comes from observations which are almost always created in a particular time/space setting such that exact replication as required in the frequentist framework becomes impossible, rendering interpretation of for example Type I error rates close to meaningless. NHST is based on a model about the long-run that in essence claim that we do know what will happen, at least in a general structural outline. But the ever-changing settings studied by social scientists ensure that long-run interpretations under NHST are almost never a directly relevant validity criterion (Greenland & Poole 2013). As Greenland and Poole (2013 p. 74) state, except for death and taxes, we almost never know what the long-run holds in store for us, it therefore makes much more sense to assume that observational data are unique and fixed at this point in time. In reality therefore, inferences from observational studies are very often based on single non-replicable results which at the same time no doubt also contain other biases besides potential sampling bias. In this respect, frequentist analyses of observational data seems to depend on unlikely assumptions that too often turn out to be so wrong as to deliver unreliable inferences, and hairsplitting interpretations of *p* values becomes even more problematic (Greenland 1990; Greenland & Poole 2013). Indeed, the low replicability of *p* values and the general failure of prediction in the social sciences warrant such a claim (see e.g., Schrodt 2006; Starbuck 2006; Taagepera 2008; Armstrong 2012).

**Some persistent misinterpretations of NHST**

It is indeed a messy situation from where confusions, misinterpretations and misuse have flourished in the social and behavioral sciences. The pathologies that emerge are detrimental, besides the mix-up of *p* values and Type I error rates, there are confusions over the meaning of 'statistical significance', confusions over the order of the conditional probability and confusions about the probability of rejection, there are logical inconsistencies coming from the probabilistic use of *modus tollens* for inference, as well as adverse behaviors of chasing 'significance' but ignoring effect size and adherence to the completely arbitrary significance thresholds, to name some of the most persistent problems; for a more exhaustive treatment we refer to Kline (2004) and Goodman (2008), who list a catalogue of major and minor misinterpretations and criticisms. Below, we discuss a few of these pathologies and in the next section we exemplify them based on two scientometric studies.

*Confusion over the interpretation of 'statistical significance'*

As outlined above, Fisher and Neyman-Pearson had very different conceptions of 'statistical significance' and the mix-up of *p* values and Type I error rates within the NHST-framework have clearly led to a general and widespread confusion over the meaning and interpretation of 'statistical significance' (Hubbard & Bayarri 2003). However, the misinterpretation is more prevalent and endemic, because whatever 'statistical significance' may mean in a technical sense, too often such a status is equated with a theoretical or practical importance of a finding, or simply that the effect found is a genuine and replicable one (Boring 1919; Ludwig 2005; Gelman & Sterne 2006; Ziliak & McCloskey 2008). NHST do not per se measure the importance of a result (Kline 2004). As it is, NHST only addresses sampling error assuming no other errors are present in the study; yet other factors are at least as important or more important for determining the real 'significance' of findings. The significance of a finding, in its true sense, depends upon the size of the effect found and can only be evaluated subjectively in the context of research design, theory, former research, practical application and whether the result can be replicated, as indeed Fisher himself argued (e.g., Kirk 1996; Fisher 1956). A major challenge for bibliometrics, scientometrics and research evaluation is that we generally have vague or no theories to help us interpret the importance of findings; the field is to a large extend instrumental. Focusing on *p* values leads to a practice where everything that turns out to be statistically significant is treated and reported as important and publishable (Scarr 1997). But large as well as small effects can be important. It is the researcher's responsibility to explain why the observed difference or association has important consequences worthy of emphasis. For various reasons, as discussed above, *p* values are flawed evidential measures. Lower *p* values do not necessarily indicate larger effect sizes and thus more 'significant' results. It cannot be as the outcome of a significance tests is determined by at least eight factors: 1) the effect size, 2) the 'stopping rule', 3) the sample size, 4) variation among cases, 5) the complexity of the analysis (degrees of freedom), 6) the appropriateness of the statistical measures and tests used, 7) the hypothesis tested and 8) the significance level chosen (Schneider & Darcy 1984; Cohen 1990). This gives plenty of room for chasing 'significance' through non-evidential factors, and one unfortunate consequence is to neglect reporting estimated effect sizes in published research, statistically significant or not, leaving readers uninformed. Scientific advance requires an understanding of how much (Tukey 1991). Indeed, an early criticism of NHST was its sensitivity to sample size (Berkson 1938; 1942). As sample size approaches infinity, $H_0$ will always be rejected because no model is accurate to a large number of decimal places (e.g., Mayo 1996). So when samples can be obtained relatively easy, like many samples of bibliometric data, large sample sizes can lead to detection of numerous trivial but 'significant' findings.

NHST is computed based on the assumption that $H_0$ is a true parameter in the population, but several critics have pointed out that nil null hypotheses are most often statements already known to be implausible to begin with in non-experimental studies (e.g., Berkson 1942; Lykken 1968; Meehl 1978; Webster & Starbuck 1988; Cohen 1994). We claim that this is also the situation in numerous scientometric studies. Many researchers will probably argue that $H_0$ is just a 'straw man' and that they are perfectly aware that $H_0$ is most likely false to begin with. But if one assumes that $H_0$ is false to begin with, the practice of testing the null becomes uninformative and resulting *p* values meaningless as they are calculated conditional on $H_0$ being true. The results tell us little except whether our sample size was sufficient to detect the difference. In Neyman-Pearson's model, Type I errors would be irrelevant if $H_0$ is false to begin with. A much more sophisticated possibility is to define a range of effect sizes that is sufficiently close to zero to represent a 'null effect'. The advantage of incorporating some notion of effect size is that it moves the researcher's attention from the question of 'statistical significance' to the more important topic of substantive significance of his or her results. A great deal more information can be extracted from a study if the focus is on interval estimation. Here we do not rely upon implausible null hypotheses. What confidence intervals (CI) gives us is more and more precise information about uncertainty, direction and magnitude of the point estimate than the *p* value is capable of. Remember though, that a correct frequentist interpretation of a 95% CI is that over unlimited repetitions of the study, the CI will contain the true parameter 95% of the times. The probability whether the true parameter value is contained in the present CI is either zero or one, but we do not know. CIs are bounded in frequentist logic and interpretation which demands some leap of faith in non-experimental settings.

Finally, in the null ritual, binary decisions are practiced instead of inference or even better estimation of uncertainty. Within Neyman-Pearson's model, binary decisions are appropriate in relation to error control 'in the long-run', not evidence applicable to a specific study. In most research contexts, when it comes to *p* values and evidential claims about the actual results, it is not appropriate to have to make an all or nothing decision about whether to 'reject' a particular null hypothesis, as Fisher himself stressed in later writings (e.g., Fisher 1956). Thresholds for significance levels are arbitrary and in research contexts it is absurd to have to conclude one thing if the result gives $p = .051$ and the exact opposite if $p = .049$ because (Rosnow & Rosenthal 1988).

*Confusion over the order of conditional probabilities*

The literature discussing *p* values can roughly be divided into two dominant themes: 1) critique of *p* values as conceptually incoherent and essentially flawed evidential measures (e.g., Goodman 1999a; 1999b; Wagenmakers 2007), and 2) persistent and widespread problems with the interpretation and logic of *p* values (e.g., Kline 2004; Goodman 2008). In the previous section we discussed the widespread misinterpretation of linking 'statistical significance' to importance. In this section we discuss some of the other frequent misinterpretations. When interpreting *p* values one should not forget that *p* is a conditional probability, i.e., the probability of the observed data plus more extreme data, *conditional* on $H_0$ being true, that the sampling method is random, that all distributional requirements are met, that scores are independent and reliable, and that there is no source of error besides sampling error (e.g., Kline 2013). The general form can be written as $p(D+|H_0$ and all other assumptions of the model holds) for short $p(D+|H_0)$. If any of these assumptions are untenable, $p$ values will be inaccurate, often too small, and difficult to interpret (Berk & Freedman, 2003).

The most pervasive misunderstanding of $p$ values relates to confusions over the order of this conditional probability. Many regard $p$ values as a statement about the probability of a null hypothesis being true or conversely, $1 - p$ as the probability of the alternative hypothesis being true (e.g., Carver 1978; Cohen 1994; Kline 2004; Goodman 2008). But a $p$ value cannot be a statement about the probability of the truth or falsity of any hypothesis because the calculation of $p$ is based on the assumption that the null hypothesis *is* true in the population. This is known as the 'permanent illusion' (Gigerenzer, 1993) and it is widespread, also among teachers of statistics (Haller & Krauss 2002) and in textbooks (Nickerson 2000) and it has severe consequences for the interpretation of NHST. According to Schwab et al. (2011), the logic of NHST is very difficult to comprehend because it involves double negatives and often assumptions that are clearly false to begin with. Disproving the impossible (a false $H_0$) is such unusual logic that it makes many people uncomfortable, as indeed it should (Schwab et al., 2011). We posit a nil null hypothesis which is most likely implausible, and then argue that the observed data and more extreme ones would be very unlikely if $H_0$ were true. So according to Schwab et al. (2011), a finding of statistical significance states that the observed and more extreme data would be very unlikely if the impossible was true! This is nonsense, but it is also understandable that we try to inject some sense into this, where, unfortunately, the most pervasive one, is the 'permanent illusion', i.e., the fallacy of treating '$p < \alpha$' as the probability that $H_0$ is true given the data, $p(H_0|D)$. To know this, Bayes' theorem and prior probabilities are needed and this is certainly not in the frequentist toolbox (Cohen 1994). According to Cohen (1994 p. 997) "[NHST] does not tell us what we want to know, and we so much want to know what we want to know that, out of desperation, we nevertheless believe that it does! What we want to know is 'Given these data, what is the probability that $H_0$ is true?' But as most of us know, what it tells us is 'Given that $H_0$ is true, what is the probability of these (or more extreme) data?'" (i.e., $p(D|H_0)$). Stated more formally, it is a fallacy to believe that obtaining data in a region of a distribution whose conditional probability under a given hypothesis is low implies that the conditioning hypothesis itself is improbable. Cohen (1994) argues that, because of this fallacy, NHST lulls quantitative researchers into a false sense of epistemic certainty by leaving them with the "illusion of attaining improbability" (p. 998).

To complicate matters, in one sense, the 'permanent illusion' fallacy can be seen as a probabilistic variant of a classic rule of logic (*modus tollens*) (Pollard & Richardson 1987; Krämer & Gigerenzer 2005). Several authors have argued that the underlying logic of NHST suffers from severe limitations that render the information the technique generates less than definitive for judging the legitimacy of outcome generalizations (e.g., Berkson 1942; Pollard & Richardson 1987; Cohen 1994; Falk & Greenbaum 1995). In scientific reasoning, the most definitive test of a hypothesis is the syllogism of *modus tollens* or 'proof by contradiction'. With absolute statements this syllogism leads to logically correct conclusions. Consider the valid logical argument form:

> If *A* is true, then *B* is true
>
> *B* is false
>
> ∴ *A* is false

This argument is a 'proof by contradiction' as *A* is proved by 'contradicting' *B*, that is the falsehood of *A* follows from the fact that *B* is false. This is also the logical form used in NHST, however, the crucial predicament is that *modus tollens* becomes formally incorrect with probabilistic statements which may lead to seriously incorrect conclusions. The major premise and conclusion in NHST are couched in probabilistic terms as follows:

> If $H_0$ (*A*) is true, then this result is highly unlikely (*B*)
>
> This result has occurred (*B*)
>
> ∴ $H_0$ is highly unlikely (*A*).

The major premise (i.e., 'if-then' statement) leaves open the possibility that *A* may be true while *B* is nonetheless false and the conclusion may be false even if the major premises *A* and *B* are true. This is a violation of formal deductive logic which posits that the conclusion must be true when *A* and *B* are true. The problem with this approach is that it accommodates both positive and negative outcomes, so that it loses its power for enabling a researcher to evaluate any hypothesis (Pollard & Richardson 1987; Cohen 1994). Pollard and Richardson (1987) give the following example:

> If this person is American (*A*), then this person is probably not a member of Congress (*B*)
>
> This person is a member of Congress (*B*)
>
> ∴ He or she is probably not American (*A*)

This example makes plain that 'probabilistic proof by contradiction' is an illusion, it is not a valid deductive argument and yet this is literally the form of argument made by NHST. Pollard and Richardson (1987 p. 162) argue that this logically fallacy intuitively leads to a transformation from "if $H_0$ then the probability of data [a result that leads to the rejection of $H_0$] is equal to $α$" to "if data then the probability of $H_0$ is equal to $α$" as if these statements were symmetrical, they are not, the former is $p(D+|H_0)$ the latter $p(H_0|D)$. There are simply no logical reasons to doubt the genuineness of $H_0$ given that a rare event has occurred (Spielman 1974). The 'illusion of attaining improbability' undercuts the logical foundation of NHST.

The confusion between $p(D+|H_0)$ and $p(H_0|D)$ is substantial, $p(H_0|D)$ is often the most interesting from a scientific point of view, but $p(D+|H_0)$ is what NHST gives us. And this is not statistical double-talk the distinction is fundamental. It is well established that a small value of $p(D+|H_0)$, e.g., $p < .05$, can be associated with a $p(H)$ that is actually near 1, this is known as the Jeffreys-Lindley paradox (e.g., Jeffreys 1961; Lindley 1957).

Some important variants of this misinterpretation are to regard a non-significant result as confirmation or support of the null hypothesis (Kline 2004). Thus, after finding that '$p > α$', a common conclusion is something like "there is no difference". Such a conclusion concerns the actual result and is evidential, but as Fisher himself pointed out, non-significant results are inconclusive. Such a conclusion is also untenable from Neyman-Pearson's behavioral perspective. Whatever your beliefs, act as if you 'accept' the null when the test statistic is outside the rejection region, and act as if you reject it if it is in the rejection region. It is rules of behavior in the long-run and the specific results tell you nothing about how confident you should be in a hypothesis nor what strength of evidence there is for different hypotheses. Consequently, in almost all cases, failing to reject the null hypothesis implies inconclusive results. It is also important to notice the substantial difference between statistical hypotheses and research hypotheses and thus statistical and scientific inference. Too often this distinction evaporates in practice and statistical hypotheses are treated and interpreted as if they were a forthright representation of research hypotheses. They are not. Statistical hypotheses concern the behavior of observed random variables, whereas scientific hypotheses treat the phenomena of nature and man and the latter hypotheses need not have a direct connection with observed data (Clark 1963). The origin of this confusion is sometimes credited to Fisher and his lack of clarity in these matters (Hurlbert & Lombardi 2009).

Another variant is to believe that $p$ indicates the probability that a result is due to chance alone (i.e., sampling error). This is also not so, as $p$ values are calculated on the assumption that $H_0$ is true, this is the assumed 'chance model', so the probability that chance is the only explanation of the result is already taken to be 1. It is therefore illogical to view $p$ as somehow measuring the probability of chance (Carver 1978).

In practice, Fisher's *p* value is more prevalent in NHST. While Type I and Type II error rates, alternative hypotheses and statistical power are outlined in many textbook, they are conflated with Fisher's ideas, and rarely if ever in practice treated as a unified theory of 'inductive behavior;' perhaps for good reasons as discussed in a previous section. What do often appear in NHST is Type I errors (*α*) and a vague formulation of an alternative hypothesis, nowhere near the precise definition in Neyman-Pearson's model. What should be remembered is that the Type I error is also a conditional probability which can be written as *α* = *p*(reject $H_0$| $H_{0\ true}$). As discussed above, it is a pre-selected fixed measure that applies only to infinitely random selections from the same finite population, and not to an actual result in a single experiment. It is therefore mistaken to believe that *p* < .05 given *α* = .05 means that the likelihood that the decision just taken to reject $H_0$ is an 'observed Type I error' is less than 5%. This fallacy confuses the Type I error with the conditional posterior probability of a Type I error given that $H_0$ has been rejected, or $p(H_{0\ true}|\text{Reject } H_0)$. Yet, *p* values are conditional probabilities of the data, so they do not apply to any specific decision to reject $H_0$ because any particular decision to do so is either right or wrong (the probability is either 1.0 or 0). Only with sufficient replication could one determine whether a decision to reject $H_0$ in a particular study was correct. In this sense, the fallacy is related to the problem of reporting results with 'roving alphas'. If one sets *α* at 5% then the only meaningful claim from a Neyman-Pearson perspective is whether the test statistic is equal to or less than 5% or not. If the observed *p* value is say .006 and reported as *p* < .01 that would be misleading in as much as one implies that the test has a long-term error rate of 1%. The Type I error rate is 5% no matter what *p* value is calculated and all results equal to or below 5% means rejection of $H_0$.

**Two examples of confusions and misinterpretations**

To illustrate the conflation between Fisher's 'significance test' and Neyman-Pearson's 'hypothesis tests', as well as some of the widespread misunderstandings when applying NHST in practice, we provide some examples from two recent studies reported in *Scientometrics* (Sandström, 2009; Barrios et. 2013). Notice, many other studies could have been chosen, our purpose is only to exemplify what we see as prevalent misinterpretations of NHST and the confusion this leads to when interpreting the empirical literature. The issues addressed should be recognizable in many other articles within our field, though obviously to varying degrees.

When NHST is practiced in observational studies, a typical confusion occurs when a researcher rejects $H_0$ based on a *p* value (Fisher) at a preselected *α* level (Neyman-Pearson) and subsequently habitually 'accepts' (confirms) the vaguely defined alternative statistical hypothesis (appears to be Neyman-Pearson but this is by no means so and it is also not in line with Fisher); and implicitly treats smaller $p$ values as increasingly stronger evidence against $H_0$ and implicit stronger support for the alternative hypothesis (the former is Fisher's measure of evidence, the latter practice Fisher would strongly object to). The confusion is exacerbated, when researchers use 'roving alphas' (i.e., $p < .05$, $p < .01$ etc.) to indicate significance at different levels; here posterior evidence (Fisher) and a priori fixed error rates (Neyman-Pearson) are mixed-up simultaneously.

Like numerous other studies, both examples echo the 'null ritual', with its conflation of Fisher and Neyman-Pearson concepts and an inherent understanding of a 'significant' result as being genuine and important. Also, like most other studies applying NHST, the two examples do not seriously reflect upon the basic conditional assumptions required in order for standard errors and $p$ values to be meaningful.

Barrios et al. (2013) investigate possible gender inequalities in Spanish publication output in psychology. One conclusion goes like this: "... the data ... show a statistically significant difference in the proportion of female authors depending on the gender of the first author ($t = 2.707$, $df = 473$, $p = 0.007$) ... thus, when the first author was female, the average proportion of female co-authors per paper was 0.49 (SD 0.38, CI 0.53-0.45) ... whereas when a man was the first author, the average proportion of women dropped to 0.39 (SD 0.39, CI 0.43-0.35)". Contrary to many studies, test statistics, degrees of freedom, standard deviations and CIs are reported. This is laudable, but the information is unfortunately not reflected upon. The approach seems to follow Fisher, reporting exact $p$ values and interpreting them as evidence against $H_0$. But a closer look reveals conflation with Neyman-Pearson concepts, for example, a pre-selected arbitrary significance level of 5% is chosen for all analyses with resulting binary decisions and 95% CIs are also reported. Notice, the frequentist (Neyman-Pearson) interpretation is that 95% of the CIs one would draw in repeated samples will include the fixed population parameter $\theta$. Whether the actual CI contains $\theta$ is unanswerable in the frequentist conception and a major reason why Fisher also disliked CIs for scientific purposes. Consequently, in the above quotation we see the implicit use of $p$ values as measures of evidence, fixed significance levels, which may either be used in the way Fisher originally suggested, i.e. a 5% threshold, or as Neyman-Pearson error rates, $\alpha$? And finally, a CI (Neyman-Pearson) is included which requires a long-run repeated sampling interpretation which is not straightforward with non-experimental data. CIs have other virtues though, which we will address in the next section.

In the example above, the unstated statistical hypothesis, $H_0$, is no difference in the proportion of female authors given the gender of the first author, and since $p < .05$, $H_0$ is rejected and the rhetoric implies that the alternative hypothesis, which basically corresponds to the actual research hypothesis of some gender inequality, is supported and consequently a 'significant' result is found. But what is implied by "a statistically significant difference"? The most likely conjecture is that a 'significant' finding means that a genuine and important effect is presumably detected. It is important to remember, that *p* values and CIs *only* address 'random errors', assuming that other biases are absent from a study and that the statistical model used is correct. *P* values do not tell us whether an effect is present or absent, but instead only measure compatibility between the data and the model they assume, including the test hypothesis. Therefore, as discussed above, *p* values are also not probabilities of results being due to chance alone (random errors). Such a probability is already taken to be 1 since the test assumes that every assumption used is correct, thus leaving chance alone to produce the difference observed. Hence, it cannot be a statement about whether these assumptions are correct but a statement what would follow logically if these assumptions *were* correct. Obviously, such conditional information is much more restricted than usually acknowledged. Close to no one interpret their *p* values or error rates as the conditional probabilities they are. For the sake of argument, let us assume that assumptions are true in the above analysis, and the result, being 'significant', is *probably* 'genuine' but that in itself does not make the result important. To judge importance, the effect size and its potential theoretical and practical importance needs to be considered. In the present example (Barrios et al. 2013), the actual effect size and its potential importance is not discussed, i.e., the average proportion of female co-authors per paper conditioned on the gender of the first author. For want of something better to compare with, we calculated a standardized effect size for the difference between female co-authors conditioned on the gender of the first author and it corresponds to a 'small effect' according to Cohen's traditional benchmarks (Cohen 1988). If we accept this benchmark as sound, then it challenges the study's claim of a 'significantly higher' effect when the first author was female (Barrios et al. 2013, p.15), and it certainly raises important semantic and epistemic questions of what precisely 'significantly higher' imply? Notice, that a *t* test with $df = 473$ has considerable power and will be able to detect small differences (Cohen 1988).

Clearly, the authors invest an epistemic value into the calculated *p* value (Fisher). But like most others, the authors oversell the epistemic value because what the *p* value says in this example is: the probability of the observed *t*-statistic or more extreme unobserved *t*-statistics *if* the statistical model used to compute the *p* value is correct. Again for the sake of argument, if assumptions are indeed correct, so we are told that the *t*-statistic is highly unlikely under $H_0$ but, as discussed in the previous sections, this is not the same as $H_0$ is implausible. In fact, gender research in general suggests that some difference is to be expected and not the other way around. But a priori knowledge rarely influences null hypothesis formulations. Also, that the *t*-statistic is highly unlikely under $H_0$ corresponds to $p(D+|H_0)$, but the conclusion that the difference is very likely, or $H_0$ is very unlikely, corresponds to $p(H_0|D)$ and such an inference is the 'inverse fallacy' (Klein, 2004). A correct interpretation is more restricted and needs to emphasize *p* as a conditional probability of the data under $H_0$.

Finally, in another test the authors conclude that: "… the data did not show a statistically significant relationship between the proportion of female authors and the number of citations received, controlled by the number of authors who signed each paper and the journal impact factor ($r_{AB.CD}$ = -0.085, *p* = 0.052) (Barrios et al. 2013, p.19). Such a statement clearly demonstrates the mindless use of NHST, why would *p* = .049 be significant and as it is *p* = .052 not? Clearly, the difference has no real life implication. If the authors were interested in Neyman-Pearson's error rates, however, the hairsplitting decision based on a preselected fixed *α* level at 5% would give meaning. But there is nothing indicating that the study comply with frequentist long-run 'inductive behaviors' and the results are certainly not interpreted as such. Instead, *p* values are used in Fisher's conception as evidential measures and in that respect it is close to absurd to declare *p* = .052 as not significant when *p* = .049 would have resulted in the opposite conclusion. Obviously, the correlation coefficient is miniscule, nevertheless, declaring the relation not statistically significant implies that the *p* value is used to decide the importance of the result – what would the authors conclude if *r* = .085 and *p* = .049?

The second example also concerns the interpretation of non-significant results. It is a study by Sandström (2009) where the interest is in the relationship between funding and research output in a Swedish context also with a special interest in gender differences. Traditional OLS-regression models are pursued, *n* = 151 and 12 models are 'tested'. The unstated null hypotheses are no relationship (i.e. zero slope) and results are reported as 'roving alphas' using asterisks to indicate whether input variables are 'significant' or not. Results are basically treated in accordance with the null ritual, where statistical significant variables are treated as important per se, the size of effects are not discussed. However, according to the author "a surprising result is that the share of basic, strategic or user-need funding does not seem to produce any differences in output variables. Not even broadness … has any significant effect on research output" (Sandström 2009, p. 348). Clearly the author's research hypothesis, or expectation, is that these funding variables are related to output. Nevertheless, 'non-significant relations' leads the author to claim that there are no relationships.

This interpretation is an unequivocal mix-up of Fisher and Neyman-Pearson and in this respect the claim is also mistaken. By the absence of a significant result, the author seems content in supporting $H_0$, that there is no relationship between the funding variables and output. According to Fisher's 'significance test', a non-significant result establishes no such thing. To Fisher, failure to reject $H_0$ simply means insufficient evidence against it and that this 'experiment' has failed to produce a significant relationship between funding and output, however, nowhere would this preclude there being one, as indeed the coefficients in the example indicate. The author's interpretation seems more in line with Neyman-Pearson's 'reject' or 'accept' reasoning but there is nothing in the design which indicates that the observational study adheres to the Neyman-Pearson doctrine in which 'hypothesis testing' must be interpreted. The conclusion of no relationship is presented by the author as a scientifically credible finding. But as outlined above, Neyman-Pearson's doctrine is a *decision* between two competing hypotheses given the power of the test, and where 'accept' has nothing to do with a scientific claim of no relation of H, it is a *decision* based on long-run rules of behavior. A non-significant result can only have these two interpretations, whereas a supposed scientific claim of no difference or no relation is erroneous.

We think that many scientometricians will be familiar with the above description of NHST practice, where a 'significant' result is treated as an important and reliable finding and a 'non-significant' result the opposite. We think it is of fundamental importance to be able to distinguish between Fisher' and Neyman-Pearson's different ideas for an acute appreciation of the modern hybrid of NHST. Granted, definitions are abstract and not intuitive and some of the logic is flawed. But this only emphasizes the underlying problems of NHST and the urgent need to be cautious when interpreting results based upon it. The final section points to some alternatives and suggests some guidance for reporting and interpreting statistical results in scientometric studies.

## Alternatives and some suggested guidance for reporting statistical results in scientometric studies

There are no easy fixes or magical alternatives to NHST (Cohen 1990). In fact, we could argue that there is nothing to be replaced because NHST does not provide what we think it provides in relation to hypothesis testing (Cohen 1994). From our point of view the best solution is to simply stop using

NHST as practiced, at least with non-experimental data. Some have even suggested a ban (e.g., Hunter 1997) but this is not the way forward – but neither is status quo. What we need is statistical reforms (e.g., Cumming 2012).

If NHST is used for research and assessment purposes, it should be in an educated and judicious way where its influence is restricted to its rightful epistemic level, which is close to nothing, as both Fisher and Neyman-Pearson acknowledged (Fisher 1956; Neyman & Pearson 1993a). In some fields like psychology and medicine, where the debate has been ongoing for decades, general guidelines for reporting statistical results exist (e.g., APA 2010) and some journals even have their own stricter rules[4]. Guidelines in themselves can be problematic, but in the case of NHST we think some guidance is urgently needed for our field. We provide some suggestions below, to inspiration for authors, reviewers and editors.

First of all, there are inferential alternatives, which contrary to NHST do in fact assess the degree of support that data provide for hypotheses, e.g., Bayesian inference (e.g., Gelman et al. 2004), model-based inference based on information theory (e.g., Anderson 2008) and likelihood inference (e.g., Royall 1997). In many ways, these alternatives lead to a greater understanding and improved inference than that provided by $p$ values and the associated statement of statistical significance. These important alternatives are unfortunately not readily available in commercial statistical software.

In all statistical analyses, focus should be on scrutinizing data. The importance of results is to be found in the data and not a mechanical decision tool. Simple, flexible, informal and largely graphical techniques of exploratory data analysis, aim to enable data to be interpreted without statistical tests of any kind (e.g., Tukey 1977).

If we stick to the frequentist philosophy, statistical reformers argue that parameter estimation should be paramount (e.g., Kline 2004; Cumming 2012), as Neyman himself preferred (Neyman 1937). Once researchers recognize that most of their research questions are really ones of parameter estimation, the appeal of NHST will wane. It is argued that researchers will find it much more important to report estimates of effect sizes with CIs and to discuss in greater detail the sampling process and perhaps even other possible biases such as measurement errors.

If we want to address the problem of sampling error the solution is obvious: use large samples. The need for NHST primarily exists in low power situations. If one increases power, sampling error decreases, and the need for NHST diminishes. In many instances, large samples are certainly viable with bibliometric data. Notice also, that bibliometric databases enable analyses of 'apparent populations' (Berk et al. 1995). In that case, NHST becomes superfluous as there is no random sampling error. Otherwise, replications are a superior way to deal with possible sampling error. Only by demonstrating it repeatedly can we guarantee that a particular phenomenon is a reliable finding and not just an artifact of sampling. Notice, *p* values tells you nothing about the reliability of a specific result (Kruschke 2010). Study replication is also advantageous for knowledge accumulation and supports meta-analyses. Needless to say, we agree with Glass (2006), that classical inferential statistics should play a little or no role in meta-analyses.

---

[4] For example, the instructions to authors in the journal Epidemiology reads "We strongly discourage the use of P-values and language referring to statistical significance" (http://edmgr.ovid.com/epid/accounts/ifauth.htm).

CIs are often promoted as alternatives or supplements to NHST. They do provide more information and are superior to NHST and should as such be preferred in relation to parameter estimation. But CIs are not a panacea for the problems outlined in this article. They are based on the same frequentist foundation as NHST and can easily be used as a covert null ritual. It is important to emphasize that supporters of frequentist statistical inference are also dissatisfied with NHST and is mix up of Fisher and Neyman-Pearson's ideas. Some see Fisher as the villain (e.g., Ziliak & McCloskey 2008), whereas others argue that Fisher's ideas are the only ones applicable to scientific practice (e.g., Hurlbert & Lombardi 2009). Indeed, this was Fisher's own reasoning and we think it has some merit. Scientific settings suitable for Neyman-Pearson's model seem restricted. Based on the aforementioned sources on statistical reform, here are some recommendations for best practice in quantitative data analysis:

- Statistical inference only makes sense when data come from a probability sample and/or have been randomly assigned to treatment and control groups. If assumed, a stochastic mechanism should always be reflected upon in a study. Many scientometric situations seem unsuitable for the frequentist logic of inference.
- Whenever possible take an estimation framework, starting with the formulation of research aims such as "how much?" or "to what extent?" Size matters in research, ordinal relationships are for most of them trivial.
- Interpretation of research results should be based on point and interval estimates. Attention is thus given to uncertainty and sample size.
- Calculate effect size estimates and CIs to answer those questions, then interpret results based on theory, context, cost benefit, former research etc. The importance of a result is eventually an informed subjective judgment.

- If NHST is used, (a) information on statistical power or at most sample size must be reported, and (b) $H_0$ should be plausible. Do not test $H_0$ when it is clearly known to false.
- Effect sizes and CIs must be reported whenever possible for all effects studied, whether large or small, statistically significant or not. This supports knowledge accumulation and meta-analysis.
- Exact *p* values should always be reported, not *p* < at some conventional *α*. Stop using hairsplitting significance levels, 'roving alphas' and asterisk.
- It is totally unacceptable to describe results solely in terms of 'statistical significance', as if they were important.
- It is the researcher's responsibility to explain why the results have 'substantive significance' statistical tests is inadequate for this purpose.

Consider the advice given by Meehl (1990), always ask oneself, if there was no sampling error present (i.e., if these sample statistics *were* the population parameters), what would these data mean. If one feels uncomfortable confronting this question, then one is relying on significance tests for the wrong reasons. If one can answer this question confidently, then the use of significance tests will probably do you little harm, but significance tests will probably do you no good either.

NHST is, in most instances, "a very inappropriate tool used in very inappropriate ways, to achieve a misinterpreted result" (Beninger et al. 2012, p. 101). NHST is not the perceived objective procedure leading to truthful inferences. No such statistical tool exists. Even if NHST is used and understood properly, the results are usually not very informative for making inferences. NHST is poorly suited for this because it poses the wrong question, $p(D|H_0)$. Unfortunately, at the same time, we as researchers think it provides us with the correct answers, $p(H_0|D)$. This is detrimental to cumulative scientometric research.